\RequirePackage{lineno}
\documentclass[showpacs,preprintnumbers,amsmath,amssymb]{revtex4}
\usepackage{natbib}
\usepackage{graphicx}
\usepackage{dcolumn}
\usepackage{bm}
\usepackage{color}
\usepackage{everypage}

\begin{document}

\title{Complete Chaotic Mixing in an Electro-osmotic Flow by Destabilization of Key Periodic Pathlines}

\author{R. Chabreyrie $^{1}$}
\author{C. Chandre $^{2}$}
   
\author{P. Singh $^{3}$}
  
\author{N. Aubry $^{1}$}
   
\affiliation{${1}$ Department of Mechanical Engineering, Carnegie Mellon University, Pittsburgh,  Pennsylvania 15211, USA.\\
$2$ Centre de Physique Th\'eorique, CNRS-Aix-Marseille Universit\'e, Campus de Luminy, Case 907, F-13288 Marseille Cedex 09, France.\\
$3$ Department of Mechanical Engineering, New Jersey Institute of Technology, Newark, New Jersey 07102, USA.}

\begin{abstract}
The ability to generate complete, or almost complete, chaotic mixing is of great interest in numerous applications, particularly for microfluidics. For this purpose, we propose a strategy that allows us to quickly target the parameter values at which complete mixing occurs.  The technique is applied to a time periodic, two-dimensional electro-osmotic flow with spatially and temporally varying Helmoltz-Smoluchowski slip boundary conditions. The strategy consists of following the linear stability of some key periodic pathlines in parameter space (i.e., amplitude and frequency of the forcing), particularly through the bifurcation points at which such pathlines become unstable. 
\end{abstract}

\pacs{47.51.+a, 47.52.+j, 47.61.Ne} 
\maketitle

\section{Introduction}

As it is well-known, the generation of efficient and complete mixing is particularly challenging for Stokes flows, for which the Reynolds number is very low ($Re<1$) and turbulence does not exist. These include large scale flows for fluids of high and/or low velocity as well as small scale flows \cite{OttinoWiggins:2004}. At miniature scale, efficient mixing leads to important applications, particularly lab-on-chip devices that can incorporate one or several laboratory functions on a small surface (with a size of approximately one millimeter square or less). In practice, such devices are often used as biochemical reactors for which complete mixing is essential in order to bring different reagents in contact of one another and trigger chemical reactions between them.

For the past three decades, the kinematics viewpoint of fluid mixing, that is the behavior of trajectories of passive or advected particles, has received much attention (see \cite{Aref:2002}). This is through this viewpoint that the intimate relation between fluid mixing and chaotic advection, also referred to as chaotic mixing, has been explored in detail. Specifically, it is now well-understood that chaotic mixing occurs through successive mechanisms of stretching and folding of material lines and that it increases exponentially the contact area between the different fluids or reagents to be mixed.

The generation of chaotic advection in a small Reynolds number flow, particularly in a microfluidic device, is generally achieved by adding a degree of freedom to a two-dimensional incompressible base flow. Such a degree of freedom 
takes the form of time dependence \cite{Solomon:1988a, Solomon:1988b, Paoletti:2006, Mancho:2006} or  a third dimension \cite{Baj, KroujilineandStone:1999}.

Generally, this time dependence or  third dimension is periodically introduced into the system via passive techniques based on altering the device geometry \cite{Liu:2000, Stroock:2002, Mensing:2004},  active techniques based on forcing (e.g., \cite{Oddy:2001,Bau:2001, Ouldelmoctar:2003, GlasgowAubry:2003, Glasgow:2004, Friend:2008}), or the combination of both \cite{Goullet:2006,Niu:2003, Bottausci:2004,   Stremler:2004}.
In this work, we use the same base flow as in Ref. \cite{Bau:2001} with a time periodic dependence introduced into the system via forcing which is created by temporally varying  the slip boundary conditions along the channel walls. 
While the forcing was created by using a discontinuous function in the form of a switch in Ref. \cite{Bau:2001}, our forcing is created by a smooth time-periodic oscillation of the slip boundary conditions. The smoothness of the system is easier to handle computationally and experimentally.

In this paper, we focus on creating and spreading chaotic mixing within the entire channel. Our goal heads towards  the opposite direction of other works that target the total elimination of chaotic mixing  or those that target partial mixing at a desired location and size in phase space \cite{ChabreyriePRE08}.
Our strategy follows the one in Ref.~\cite{Tounsia:2007}, where the global extension of chaotic mixing can be summed up by the linear stability of a few key periodic pathlines. With this strategy complete or at least almost complete chaotic mixing is achieved. 

The paper is organized as follows.  The physical model as well as its assumptions and corresponding dynamical system are described in Sec.~\ref{Sec:Phys_Model}.
Section \ref{Sec:Characteristics} shows the complex relationship between the system parameters (i.e., amplitude and frequency of the perturbation)  and  the mixing characteristics.
In particular, this section outlines how the trial and error methods commonly used to find complete chaotic mixing is burdensome and computationally demanding. Indeed, for only one chosen pair of parameter values, such an approach (based on Poincar\'e sections, finite time Lyapunov exponent maps and mixing indices) requires the integration of the dynamical system for a large number of initial conditions over long periods of time.
In Sec. \ref{Sec:Control}, we present a strategy capable of determining the set of parameter values leading to complete, or almost complete, chaotic mixing in a much more efficient and less time consuming fashion. In contrast to the previous methods our method focuses on a few invariant structures, i.e., key periodic pathlines over a very short period of time. It allows us to identify the parameter values for which well spread chaotic mixing takes place. 

\section{\label{Sec:Phys_Model}Physical model}
\subsection{Assumptions}
\begin{figure}
\begin{center}
\includegraphics*[width=14cm]{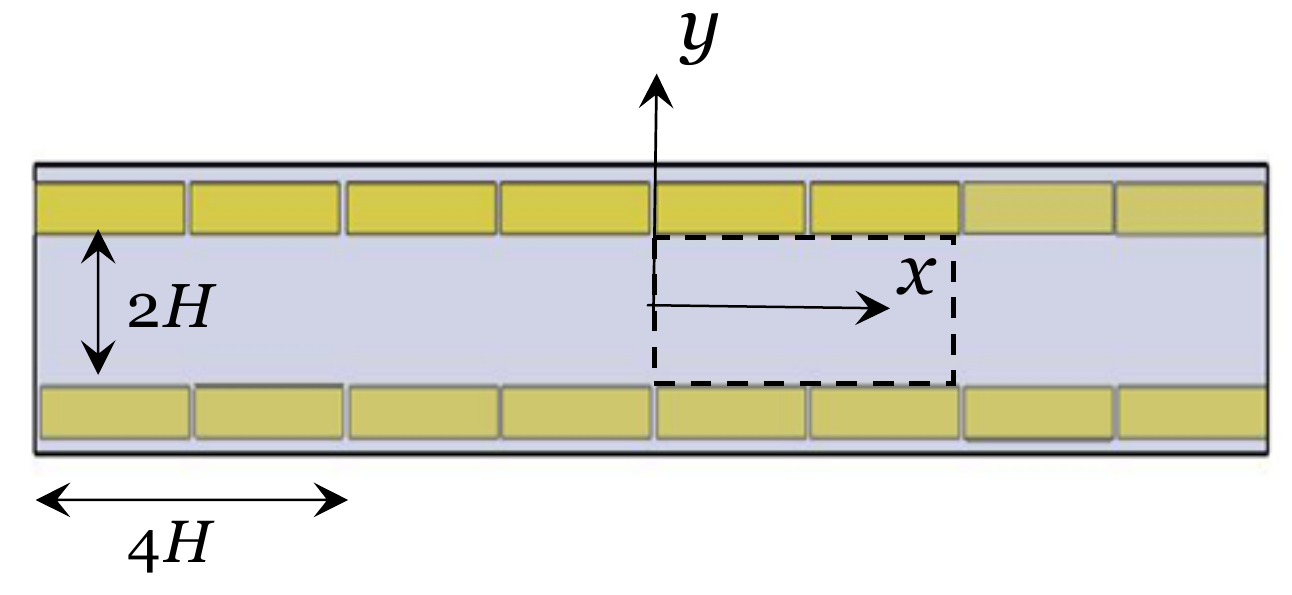}
\end{center}
\caption{Schematic of the microfluidic channel subjected to a steady uniform electric field $E_0$ in the downstream direction. In addition, the channel is equipped with a series of electrodes embedded within its walls.  The channel can thus be decomposed into a series of generic periodic cells defined by a set of four adjacent electrodes. The black dashed line represents the boundary of a generic cell.}
\label{figure0}
\end{figure}

We consider a straight two-dimensional microfluidic channel of height $2H$, filled with a weak electrolyte solution of electrical permittivity $\epsilon_{el}$ and kinematic viscosity $\nu$. Electroosmotic pumping of the fluid is achieved by applying a steady, uniform electric field $E_0$ in the streamwise direction by means of two electrodes placed upstream and downstream of the channel (see Fig.~\ref{figure0}). When the electrodes are energized, it is known that an electroosmotic flow is generated due to the accumulation of counterions near the charged channel walls and their downstream migration.  While the electrical double layer containing the counterions is typically thin compared to the channel height, it nevertheless drags the flow with it, thus resulting in a quasi-uniform flow velocity profile.  Assuming that the electrical double layer is indeed sufficiently small, the magnitude of the flow velocity is defined by the slip velocity $V_{HS}$ at the walls and its expression is given by the Helmoltz-Smoluchowski formula
\begin{equation}
V_{HS}=\frac{E_0\epsilon_{el}\zeta}{\nu},
\label{Helmoltz-Smoluchowski}
\end{equation}
where $\zeta$ is the potential difference across the electrical double layer, also called the $\zeta$ potential.  It is clear from the above expression that the flow velocity, in both amplitude and direction, can be controlled by adjusting the $\zeta$ potential.  As in Ref. \cite{Qian:2002}, this is performed by placing a series of adjacent electrodes of length $2H$ within the upper and lower walls \cite{Qian:2002} (see Fig.~\ref{figure0}). By varying the voltage applied to the side electrodes, the slip velocity $V_{HS}$ at the walls can be varied in time, independently of $E_0$  
\cite{Lee:1990, Hayes:1992, Schasfoort:1999, Buch:2001}. 

Due to the small scale of the microfluidic device (with a height typically of the order of $10-100$ micrometers) and the relatively weak electroosmotic velocity (of the order of $10^{-4}~{\rm m}.{\rm s}^{-1}$), the Reynolds number is typically much smaller than one.  
It is thus reasonable to model the flow inside the channel as an incompressible Stokes flow. 
The boundary conditions, which consist of a periodic slip velocity (in both space and time) at the upper and lower walls ($y=\pm H$, respectively), are  expressed as
\begin{equation}
{\bm V_{{\bm\lambda}}}\left(x,\pm H,t\right)={\bm V}_{HS}\left(\pm Ra_{2H}\left(x\right)+\epsilon\cos\omega t\right),
\label{slip_velocity}
\end{equation} 
where $Ra_{2H}\left(x\right)=4/\pi\sum_{n=1}^{\infty}\sin\left(\left(2n-1\right)x/2H\right)/\left(2n-1\right)$ denotes the periodic square wave function of period $2H$ bounded by the values $(-1,1)$ and ${\bm \lambda}=\left\{\epsilon,\omega\right\}$ stands for the amplitude and frequency of the forcing.  
The unsteady part of the slip boundary conditions is thus characterized by two parameters, the amplitude $\epsilon$ and the frequency $\omega$. As we show below, control of chaotic mixing will be achieved by adjusting these parameters.
\subsection{Velocity field: Dynamical system}
\begin{figure}
\begin{center}
\includegraphics*{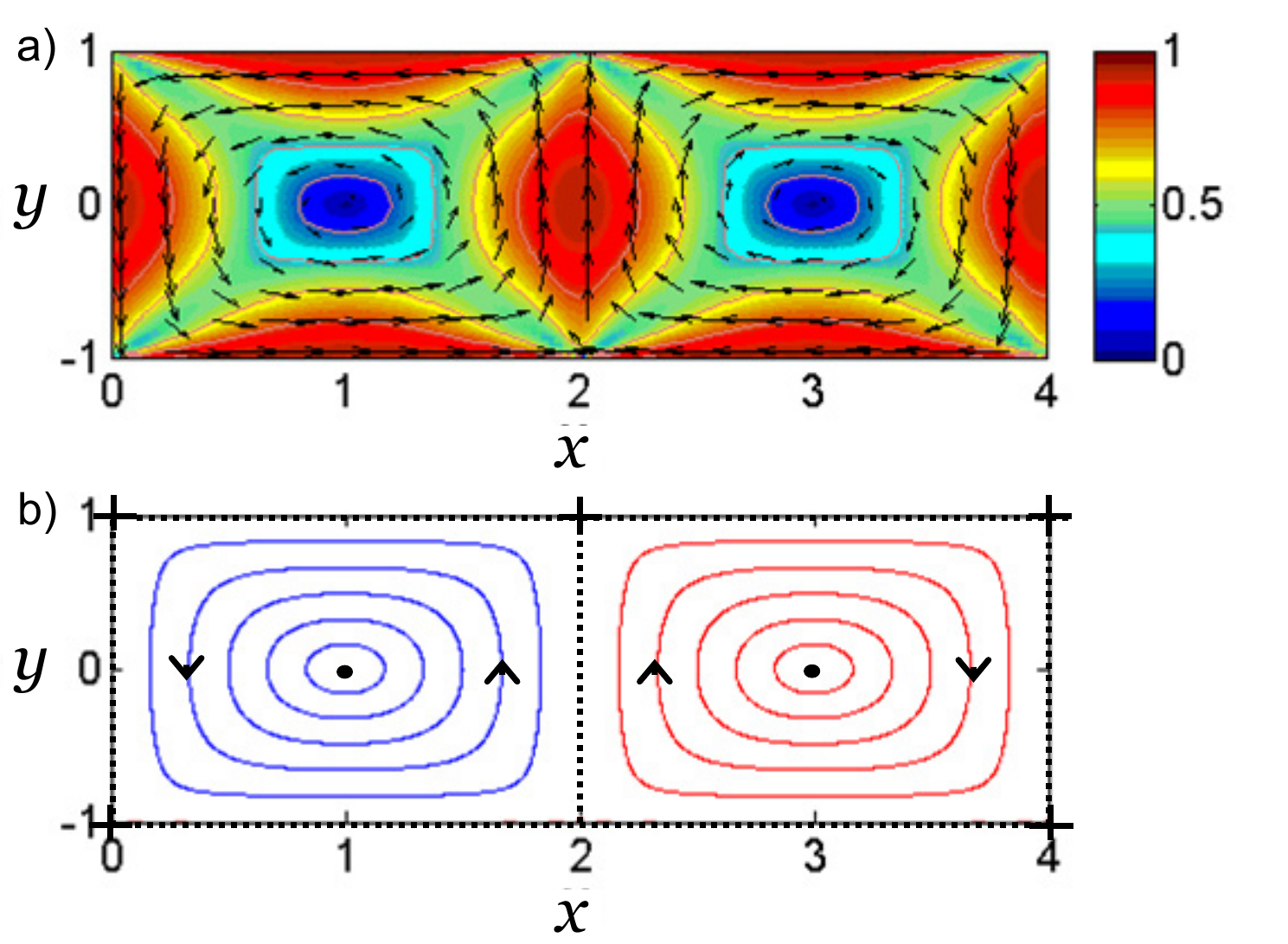}
\end{center}
\caption{a) Velocity field and contour levels of the velocity intensity inside a generic cell for the steady case ($\epsilon=0$). b) Streamlines of the flow, highlighting the heteroclinic pathlines (dashed lines), the hyperbolic fixed points (crosses) and the elliptic fixed points (thick dots).}
\label{ChapIVFig00} 
\end{figure}

Due to the linearity of the Stokes flow problem, the velocity field $\bm{V_{\lambda}}$ can be decomposed into a steady component $\bm{V}_0$ obtained for $\epsilon=0$ (see Fig.~\ref{ChapIVFig00}-a), and an unsteady one, $\bm{V}_{1}$, so that 
 \begin{equation} \bm{V_{\lambda}}\left(x,y,t\right)=\bm{V}_0\left(x,y\right)+\bm{V}_{1}\left(x,y,t\right),
 \label{flow}
 \end{equation}
where the variables have been made dimensionless by using $H$, $V_{HS}$ and $H/V_{HS}$ as length, velocity and time scales.
In the steady case, one obtains recirculating rolls in between the upper and lower walls (see Fig.~\ref{ChapIVFig00}-a), as shown by \cite{Qian:2002}. The unsteady, perturbed case is then generated by means of the unsteady slip velocity obtained by temporally varying the $\zeta$ potential at the walls.

\begin{figure}
\begin{center}
\includegraphics*{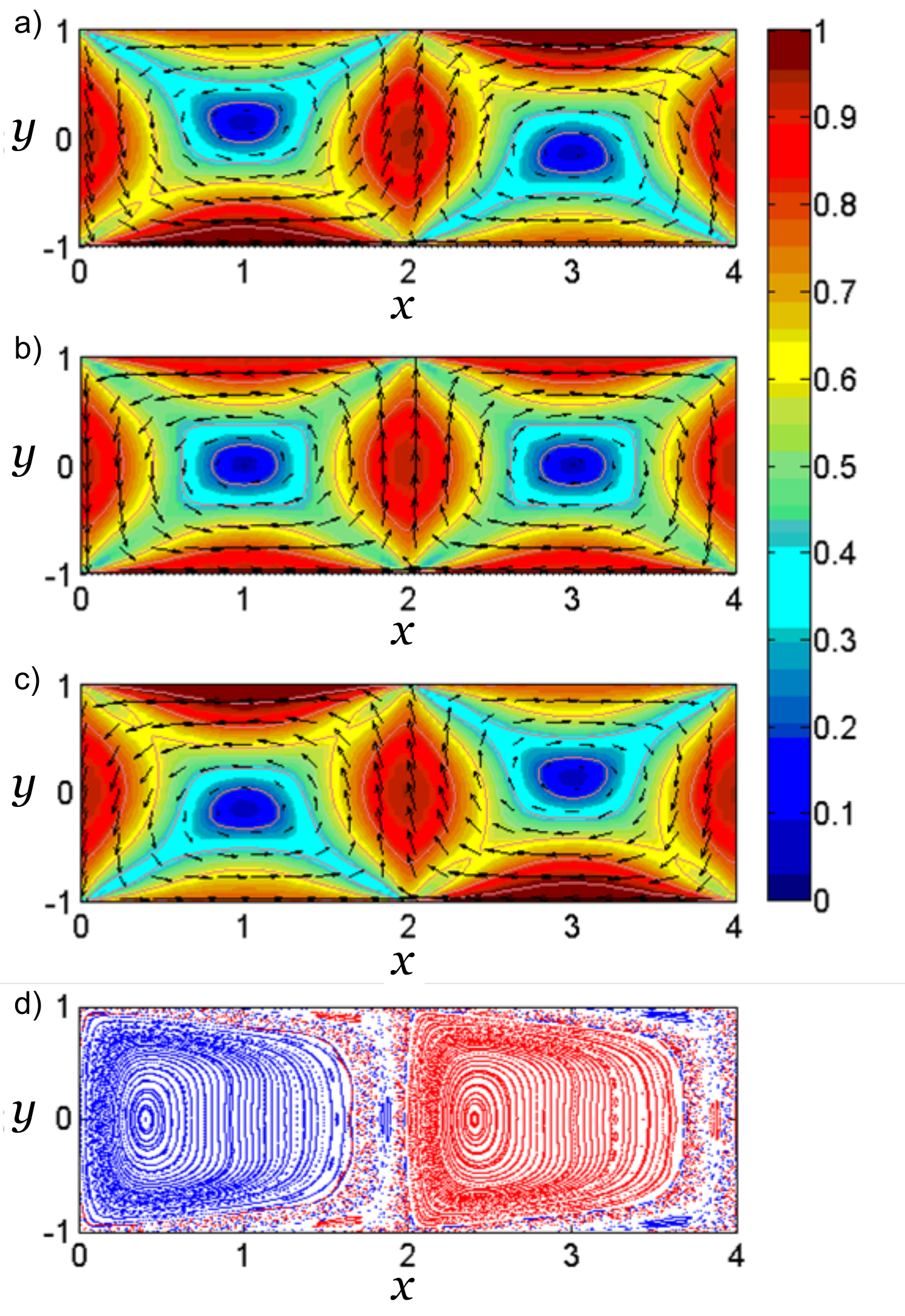}
\end{center}
\caption{a-c) Velocity field and contour levels of the velocity intensity inside a generic cell for the unsteady case with $\epsilon=0.10$ and $\omega=0.90$ at time $t=\pi/2\omega,\pi\omega,3\pi/4\omega$. d) Corresponding Poincar\'e section.}
\label{ChapIVFig3a} 
\end{figure}

For an incompressible flow, the dynamics of a passive particle can be studied by using the stream function formulation. Its velocity is expressed as
\begin{eqnarray}
\label{ham1}
u=\dot{x}&=&-\frac{\partial\psi_{{\bm \lambda}}\left(x,y,t\right)}{\partial y},\\
\label{ham2}
v=\dot{y}&=&\frac{\partial\psi_{{\bm\lambda}}\left(x,y,t\right)}{\partial x},
\end{eqnarray}
with 
\begin{equation}
\psi_{{\bm \lambda}}\left(x,y,t\right)=\psi_0\left(x,y\right)+\psi_{1}\left(x,y,t\right).
\label{stream_function}
\end{equation}
This system is Hamiltonian, with the stream function $\psi_{{\bm\lambda}}$ acting as the time dependent Hamiltonian and the physical coordinates $(x,y)$ playing the role of conjugate variables.
The stream function is composed of a steady component, $\psi_0$ (corresponding to $\bm {V}_0$), and an unsteady one, $\psi_{1}$  (corresponding to $\bm{V}_{1}$).
The steady part is determined by solving the steady Stokes equation with periodic slip velocity at the walls.
Such a problem can also be approached by means of the stream function formulation, i.e., by solving the biharmonic equation
\begin{equation}
\label{biharmo}
\nabla^2\psi_0=0,
\end{equation}
where the stream function of the steady case, $\psi_0$, satisfies impermeable and periodic slip conditions at the walls
\begin{eqnarray}
\label{impermeable_conditions}
\psi_0\left(x,\pm1\right)&=&0,\\
\label{slip_conditions}
\frac{\partial \psi_0}{\partial y}\left(x,\pm1\right)&=&\pm Ra_{2H}\left(x\right).
\end{eqnarray}
The biharmonic equation~(\ref{biharmo}) is then solved by the method of
separation of variables, the solution taking the form
\begin{equation}
\psi_0\left(x,y\right)=\sum^{\infty}_{n=1}\alpha_n X_n\left(x\right)Y_n\left(y\right)\\
\label{stream_reg_El_0},
\end{equation}
where
\begin{eqnarray}
\label{stream_reg_El_1}
X_n\left(x\right)&=&\sin\left(\frac{n\pi x}{2}\right),\\
\label{stream__El_2}
Y_n\left(y\right)&=&\frac{\tanh\left(\frac{\left(2n-1\right)\pi}{4}\right)\cosh\left(\frac{\left(2n-1\right)\pi y}{4}\right)-y\sinh\left(\frac{\left(2n-1\right)\pi y}{4}\right)}{\cosh\left(\frac{\left(2n-1\right)\pi}{4}\right)},\\
\label{stream_reg_El_3}
\alpha_n&=&\frac{\left(1-\cos\left(n\pi\right)\right)\cosh^2\left(\frac{n\pi}{2}\right)}{\left(\frac{n\pi}{2}\right)\left(\left(\frac{n\pi}{2}\right)+\cosh^2\left(\frac{n\pi}{2}\right)\tanh\left(\frac{n\pi}{2}\right)\right)}.
\end{eqnarray}
For further details on the derivation of the solution $\psi_0$, we refer the reader to reference \cite{Qian:2002}.\\
The unsteady component (or perturbation), generated by the oscillating slip conditions of amplitude and frequency ${\bm \lambda}=\left\{\epsilon,\omega\right\}$ at the walls, act in the $y$ direction only and can be trivially derived as
\begin{equation}
\psi_{1}\left(x,y,t\right)=\epsilon y\cos\omega t.
\label{stream_per_El}
\end{equation}
The resulting velocity field in a generic cell is represented in Fig.~\ref{ChapIVFig3a}a-c in space and time. 

\section{\label{Sec:Characteristics} Mixing characteristics as a function of parameters}
\subsection{Integrable case}
The steady case, corresponding to $\epsilon=0$ when the time dependent voltage applied to the wall electrodes is turned off (or equivalently, without time periodic slip conditions), is characterized by one invariant of the dynamics, the stream function $\psi_0\in\left[-\psi_{max},\psi_{max}\right]$ given by Eqs.~(\ref{stream_reg_El_0})-(\ref{stream_reg_El_3}).\\
In this case, the flow consists of a series of periodic cells, each cell consisting of two side-by-side recirculating rolls rotating in opposite directions as represented in Fig.~\ref{ChapIVFig00}b. Within a generic cell, pathlines are organized into two sets of  closed lines around elliptic fixed points, $\Gamma_{\pm\psi_{max}}$ located at $(1,0)$ and $(3,0)$.  In addition, heteroclinic pathlines, $\Gamma_{0}$, connect hyperbolic fixed points located at the corners of a generic half cell i.e., $(0,-1),(0,1),(2,-1),(2,1)$ (see Fig.~\ref{ChapIVFig00}-b).\\
For $\psi_{0}\neq0$ the corresponding pathlines are closed lines of constant $\psi_0$, denoted by $\Gamma_{\psi_0}$. Following the non-mixing structure of the pathlines, a small amount of dye injected into the flow evolves into a circular-like pattern without spreading throughout the channel. Clearly, no mixing is generated in this case.
\\
\subsection{Slightly perturbed case}
\label{sec:slightly}
With a time dependent perturbation introduced by the oscillatory slip conditions, the system acquires an additional degree of freedom, thus making the generation of chaotic mixing possible. 
We first consider the weakly perturbed system, that is the flow subjected to a perturbation of small amplitude ($\epsilon \ll 1$).  Figure~\ref{ChapIVFig3a}d displays a Poincar\'e section of the dynamics given by Eqs.~(\ref{ham1})-(\ref{ham2}), that is a stroboscopic map of period $2\pi/\omega$ modulo the length of the cell.  
The phase space in each generic cell consists of a small chaotic mixing region created by the destruction of the heteroclinic pathlines present in the unperturbed system. Deeper within the center of the two half cells, two symmetric sets of quasi-periodic pathlines revolve around elliptic periodic  pathlines (labeled as $O_{\psi_{max}}$ and $O_{-\psi_{max}}$). In such regular islands no mixing occurs.
\\
\subsection{Qualitative observation of the chaotic mixing}
\begin{figure}
\begin{center}
\includegraphics*{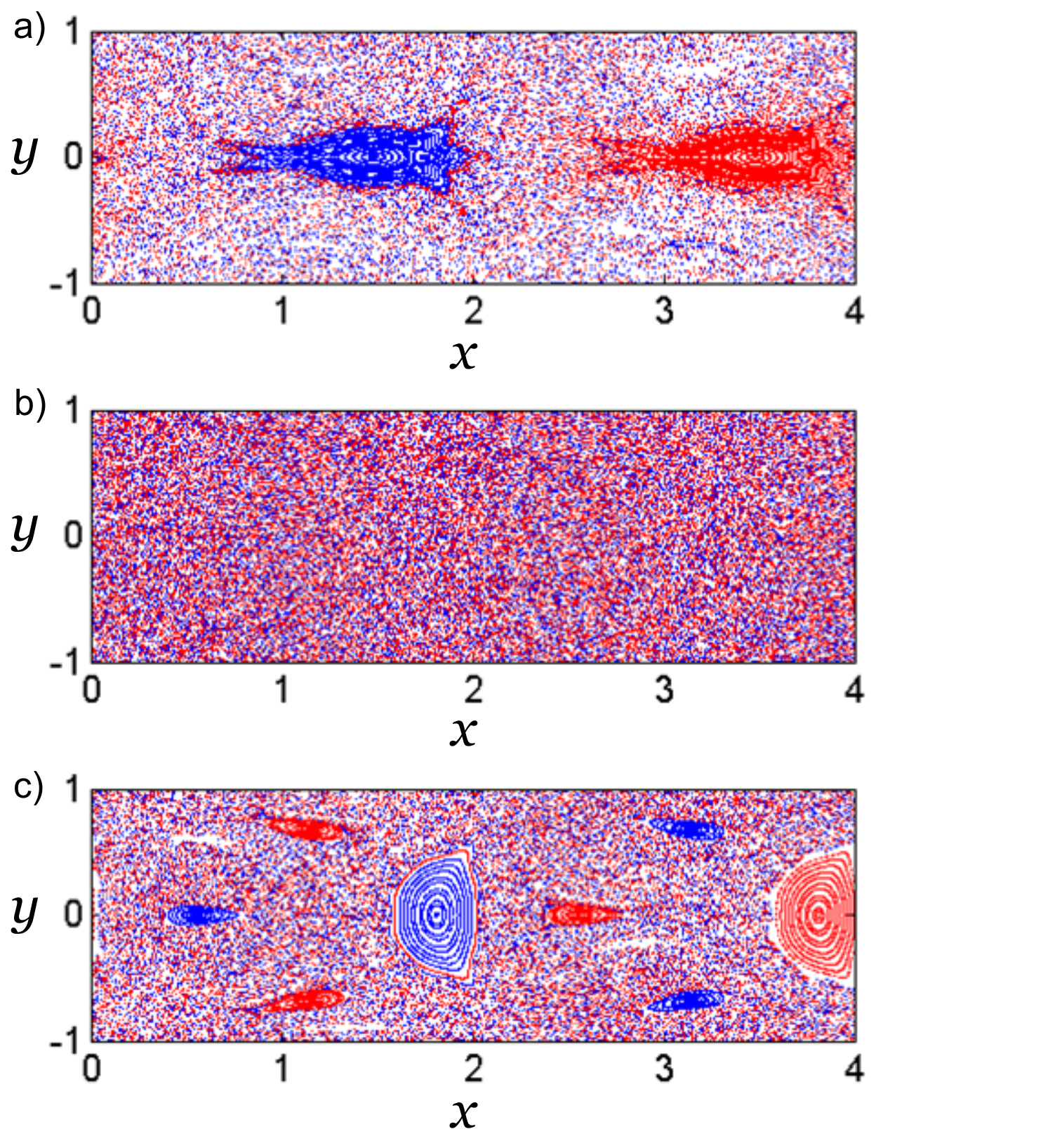}
\end{center}
\caption{Poincar\'e sections obtained for the amplitude $\epsilon= 0.975$ and the frequencies $\omega=1.500, 1.800, 2.250$ (a-c) of the perturbation for $100$ initial conditions taken on $y=0$ over an integration time $T=500 \times 2\pi/\omega$.}
\label{ChapIVFig8b}
\end{figure}

The perturbation $\psi_1$ generated by the time periodic slip conditions at the walls plays a crucial role as it is the part responsible for the creation of chaotic mixing.
Indeed, such a periodic perturbation adds a new degree of freedom to the system under the form of time $t$, thus making chaos possible. Chaotic zones in the phase space, which here corresponds to the  real space, are created by a web of higher order resonances between the 
base flow $\psi_0$ and the perturbation $\psi_1$.
The structure of  the phase space is then composed of both chaotic and non chaotic zones intertwined in a fractal structure, with a multitude of non-mixing islands as various scales, as shown by 
Poincar\'e sections (see Fig.~\ref{ChapIVFig8b}a-c). The presence of large non-mixing islands prevents the spreading of chaotic mixing throughout the entire fluid domain.

\subsection{Quantification of the intensity and spreading of chaotic mixing}
\subsubsection{Local quantification: Finite time Lyapunov exponent map}
In order to gain further insight into both the intensity and the spreading of chaotic mixing within the channel, we compute the finite-time Lyapunov exponents (FTLE).
The technique consists of associating a FTLE $\mathcal{L}$ with an initial condition ${\bm X}_{0}=(x_0,y_0,t=0)$.\\
First, we consider the time evolution of the Jacobian $ J^t\left(x,y\right)$ given by the tangent flow and the matrix of variations ${\bm \nabla}\bm{V}_{{\bm \lambda}}$ as
\begin{equation}
\frac{dJ^t}{dt}={\bm \nabla}{\bm V_{{\bm\lambda}}}\left(x,y,t\right) J^t,\\ 
\label{tangent flow}
\end{equation}
where  $J^0=I$ is the two-dimensional identity matrix. 
The FTLE map for a finite-time $t=\tau$ is then defined as 
\begin{equation}
\mathcal{L}\left({\bm X}_{0},\tau\right)=\frac{1}{\tau}\ln\left|\gamma_{max}\left({\bm X}_{0}\right)\right|,
\label{Lyaponuv_profile}
\end{equation}
where $\gamma_{max}$ is the largest  eigenvalue (in norm) of the Jacobian $J^{\tau}$.

The FTLE map ${\bm X}_0\rightarrow \mathcal{L}\left({\bm X}_0,\tau\right)$ for some given time $\tau$ allows us to distinguish between the initial conditions leading to the presence of regular (non-mixing) islands, i.e. regions  associated with small FTLEs, and the initial conditions  leading to chaotic mixing characterized by larger FTLEs. 
The structures in phase space are then easily identified and the relative sizes of the regular (non-mixing) islands determined. This tool can be used to not only determine the phase space structures but also quantify the degree of the mixing produced. Indeed, large FTLEs indicate a strong mechanism of stretching and folding, which is the archetypical mechanism of chaotic mixing. 
Due to the symmetry of the system, the domain of the FTLE map reported below has been chosen as a periodic cell of the channel, i.e. $\left\{0<x<4,-1<y<1\right\}$.

\begin{figure}
\begin{center}
\includegraphics*{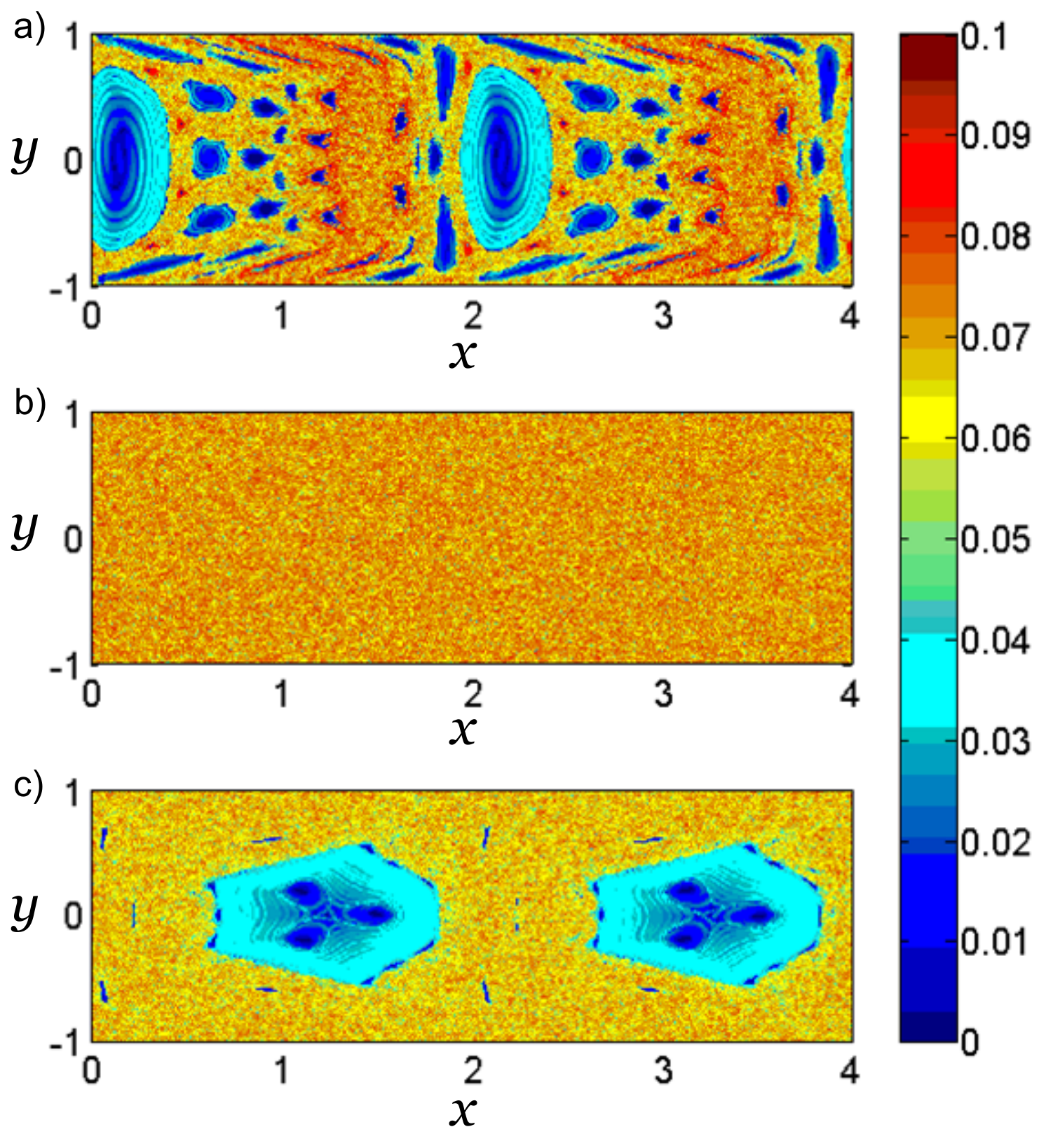}
\end{center}
\caption{Finite time Lyapunov exponent maps for $\epsilon= 0.225,0.975,1.475$ (a-c), and $\omega=1.800$. The integration time is $\tau=200\pi/\omega$ over a square grid of $400\times200$ initial conditions.}
\label{ChapIVFig6a}
\end{figure} 

Figure \ref{ChapIVFig6a} shows the FTLE map at  time $\tau=200\pi/\omega$ for the set of parameter values $\epsilon = 0.225, 0.975, 1.475$ for $\omega=1.800$. 
It is clear that in Figs.~\ref{ChapIVFig6a}a and \ref{ChapIVFig6a}c (see also Figs.~\ref{ChapIVFig8b}a and \ref{ChapIVFig8b}c) chaotic mixing is far from being complete. Indeed, around  regular islands mixing is weak (i.e., cold color), while away from these islands  mixing is very strong (i.e., hot color).
In  Fig.~\ref{ChapIVFig6a}b (see also Fig.~\ref{ChapIVFig8b}b), chaotic mixing is clearly well spread, as indicated by the more uniform FTLE map across the cell.
\\
\subsubsection{Global quantification: box counting method}
Another way to quantify the degree of mixing as a function of spatial location is by  determining the mixing index $M$  through the box counting method (see Ref.~\cite{Stremler:08}). It offers the advantage of being rather easy to implement, fast and rather cheap in computing power.
For this, we follow $N_p$ advected particles and divide the domain into $N_x \times N_y$ boxes or cells.  At each time, the number of particles $n_i$ is computed in each box, and therefore the fraction of the total number of particles, or particle rate $r_i$.  
Given the number of particles $n_i$ inside each box $i$, the computation is performed as follows.
\begin{eqnarray}
r_i&=&\frac{n_i}{n_p}~~\mbox{ if } n_i<n_p,\\
\nonumber
r_i&=&1~~~~~\mbox{ if } n_i\geq n_p,
\nonumber
\end{eqnarray}
where $n_p$ is the average number of advected particles i.e., $n_p=N_p/\left(N_x N_y\right)$.
After computing the fraction of particles in each box and at each time $t$, the time evolution of the mixing index $M\left(t\right)$ is calculated by taking the average over all the boxes, i.e.,
\begin{equation}
M\left(t\right)=\frac{1}{N_x N_y}\sum_{i=1}^{N_x N_y}r_i\left(t\right).
\nonumber
\end{equation}
A mixing index converging towards zero ($\lim_{t \to +\infty} M\left(t\right)=0$) indicates an extremely weak mixing process, while a mixing index converging towards one ($\lim_{t \to +\infty} M\left(t\right)=1$) corresponds to a perfect mixing process.

\begin{figure}
\begin{center}
\includegraphics*[width=12cm,height=10cm]{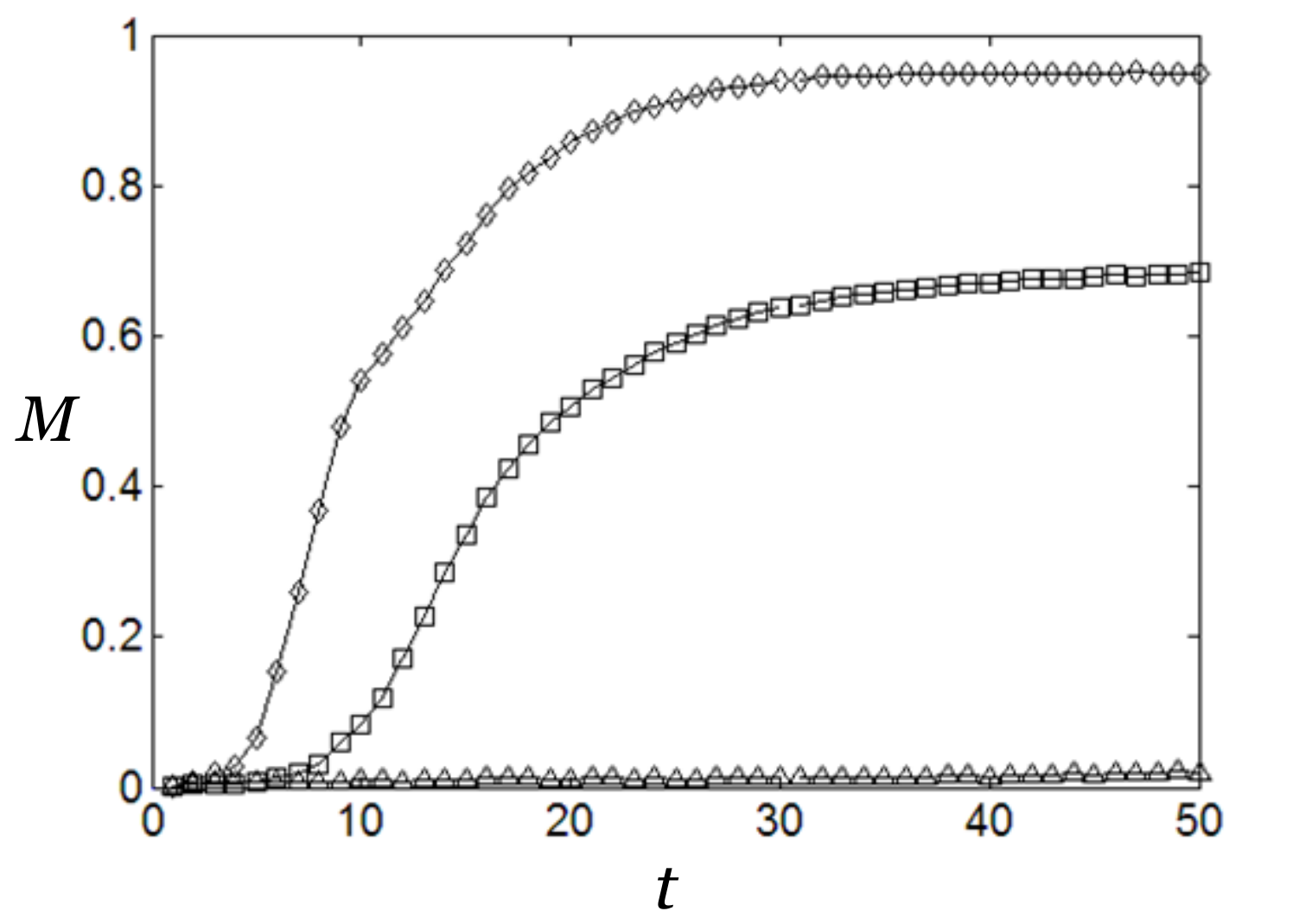}
\end{center}
\caption{Mixing index $M$ versus time $t$ in terms of the number of periods $2\pi/\omega$, computed by tracing $N_p=200,000$ advected particles over a cell domain divided into $N_x \times N_y=3200$ boxes, for the parameter values $(\epsilon=0.975;\omega=1.800)$ (diamond), $(\epsilon=0.975;\omega=2.250)$ (square) and $(\epsilon=0.975;\omega=1.500)$ (triangle).}
\label{ChapIVFig9}
\end{figure} 

Figure~\ref{ChapIVFig9} illustrates the mixing index $M$ versus the time $t$ expressed in terms of the number of periods $2\pi/\omega$, for three sets of parameter values $(\epsilon=0.975;\omega=1.800)$ (diamond), $(\epsilon=0.975;\omega=2.250)$ (square) and $(\epsilon=0.975;\omega=1.800)$ (triangle). 
In the first case, no non-mixing islands of important size are present (see Fig.~\ref{ChapIVFig8b}b), while in the other two cases one observes the existence of important non-mixing islands. From Fig.~\ref{ChapIVFig9}, it is clear that the most efficient mixing (in the  sense of a mixing index close to one) is obtained in the first case. In the former case where the chaotic mixing is well spread, Fig.~\ref{ChapIVFig9}(diamond)  shows that the mixing index $M$ quickly reaches a high asymptotic value  (after about $30$ periods),  namely $M\approx 0.95$. In Fig.~\ref{ChapIVFig9} (square), the asymptote is also reached quickly, but its value is much lower, indicating only partial chaotic mixing.
In Fig.~\ref{ChapIVFig9} (triangle), $M$ remains at a very low value, indicating poor mixing throughout the channel.

The previous trial and error methods relying on the Poincar\'e sections, FTLE maps and the mixing index to find the parameter values leading to complete mixing are extremely precise and rich in information. Unfortunately, such analyses require the integration of numerous initial conditions over long time periods for each pair of parameter values, which is extremely computationally demanding. 
In what follows we propose a much more efficient and refined strategy to determine the set of parameter values leading to well spread chaotic mixing. 
Specifically, instead of investigating numerous trajectories over long time periods
at arbitrarily chosen  parameter values, we  focus only on a few specific invariant structures of the system, e.g.,  key periodic pathlines, and track the evolution of their linear stability as a function of the parameter values.  
 
\section{\label{Sec:Control}Controlling mixing with key periodic pathlines}
\subsection{Tracking the stability of key periodic pathlines}
While a chaotic zone usually consists of a chaotic sea containing regular islands surrounding remaining elliptic pathlines, we seek to eliminate such non-mixing islands in order to obtain almost complete chaotic mixing. 

Our approach is based on the fact that a variation of the parameters, even small, is capable of influencing the dynamics in a significant manner. Our goal is then to identify the range of parameters for which complete, or almost complete, chaotic mixing occurs.  For this purpose, we study the linear stability of some key periodic pathlines in parameter space, with such pathlines being chosen from the slightly perturbed system ($\epsilon \ll 1$).

Specifically, the key periodic pathlines are selected as the twin pair of periodic pathlines $O_{\pm\psi_{max}}$ located in the middle of the two regular islands (see Fig.~\ref{ChapIVFig3a}d). These periodic pathlines come from the preservation of the strong fixed elliptic points $\Gamma_{\pm\psi_{max}}$ under slight perturbation while the pathlines near the fragile heteroclinic pathlines $\Gamma_{0}$ have been broken down to become chaotic.

\begin{figure}
\begin{center}
\includegraphics*[width=14cm]{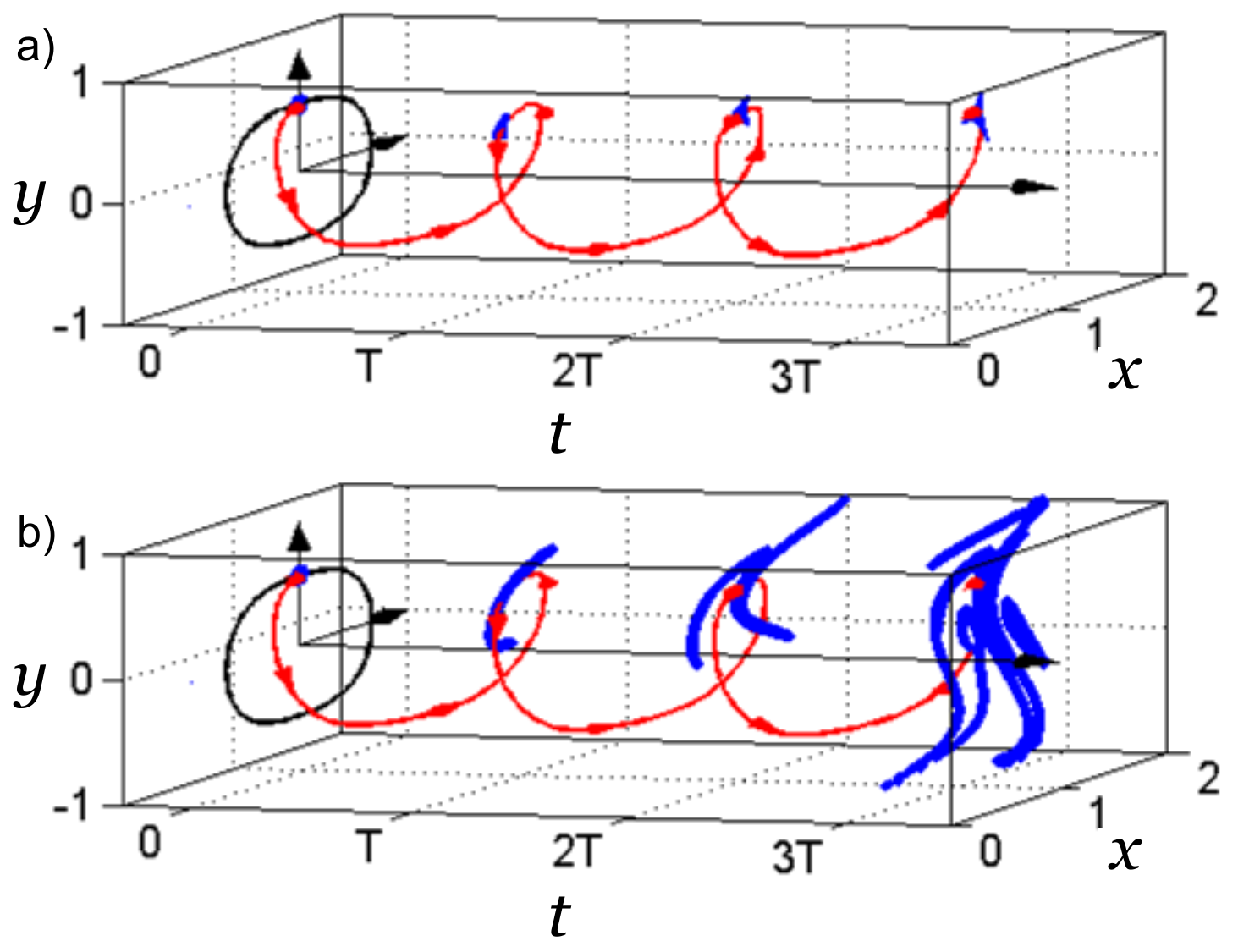}
\end{center}
\caption{Color on line. Time evolution of a periodic pathline (red) with its projection on the $(x,y)$ plane (black) and deformation of a small disc of neighboring  initial conditions (blue): a) Stable  elliptic periodic pathline obtained for the parameter values $\left(\epsilon=0.825,\omega=1.800\right)$, showing the confined deformation of the neighboring disc; b) Unstable  hyperbolic periodic pathline  obtained for the parameter values $\left(\epsilon=0.975,\omega=1.800\right)$, showing the deformation of the initial neighboring disc consisting of successive stretching and folding mechanisms.}
\label{figure_El_O_00}
\end{figure}

The stability analysis of a periodic pathline enables us to predict how a small disc of initial conditions around the periodic pathline evolves (see Fig.~\ref{figure_El_O_00}). For a stable periodic pathline the disc is slightly deformed (see Fig.~\ref{figure_El_O_00}a), while for an unstable hyperbolic periodic pathline, the disc stretches and compresses exponentially in different directions (see Fig.~\ref{figure_El_O_00}b). As it is well-known, such an unstable periodic pathline can generate chaos by an infinite number of interactions between its stable and unstable manifolds. 

The periodic pathline under investigation $O=\left\{\left(x\left(t\right),y\left(t\right)\right)\right\}_{t\in[0,T[}$ of period $T$ is numerically tracked in parameter space using a Newton-Raphson scheme (see Ref.~\cite{ChaosBook}).  In order to analyze its linear stability properties we consider the time evolution of the Jacobian $ J^t\left(x,y\right)$ given by the tangent flow (see Eq.~(\ref{tangent flow})~). 
The linear stability of the periodic pathline $O$ is then given by the spectrum of the two-dimensional monodromy matrix $J^T$, where $T$ refers to the period of the pathline. Indeed,
$J_{i,j}^T\left(\bm{X}_0\right)=\frac{\partial X_i\left(t\right)}{\partial X_j}|_{\bm{X}=\bm{X}_0}$ describes the deformation at time $t=T$ of an infinitesimal sphere of neighboring initial conditions surrounding the periodic pathline $O$ starting at $\bm{X}_0=\left(x_0,y_0,t=0\right)$.
Since the flow is incompressible, i.e.\ $\det J^T=1$, these properties can be represented by the trace of the monodromy matrix or, equivalently, by the value of the Greene's residue \cite{Greene:79,MacKay:92}~ given by 
\begin{equation}
R_{O}\left(\lambda\right)=\frac{2-{\rm tr}J^{T}}{4}.
\nonumber
\end{equation}
In particular, if $R_{O}\left(\lambda\right)\in]0,1[$ the periodic pathline is elliptic and (in general) surrounded by an elliptic (non-mixing) island;
if $R_{O}>1$ or $R_{O}<0$, the pathline is hyperbolic and thus likely to lead to the appearance of chaotic behavior around it; and if $R_{O}=0$ and $R_{O}=1$, it is parabolic.

\subsection{Destabilization of key periodic pathlines}

\begin{figure}
\begin{center}
\includegraphics*[width=12cm,height=10cm]{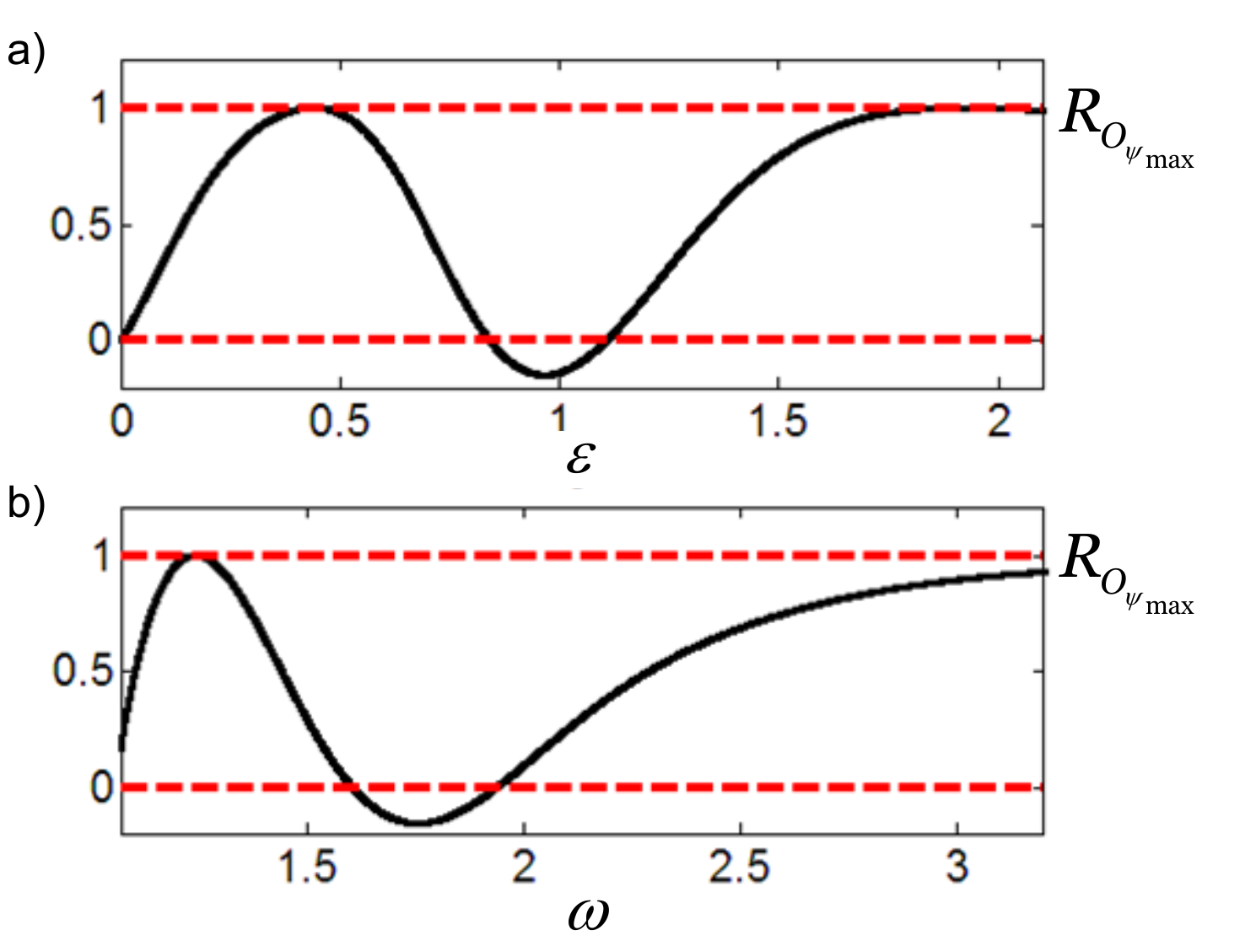}
\end{center}
\caption{Residue of the key periodic pathline $O_{\psi_{max}}$ as a function of the perturbation frequency $\omega$ and amplitude $\epsilon$ (a-b).}
\label{ChapIVFig7a}
\end{figure} 

In what follows we select the two periodic pathlines located at the center of the non-mixing regions in Fig.~\ref{ChapIVFig00}b and \ref{ChapIVFig3a}d as key periodic pathlines. They correspond to the values of the stream function $\psi_0$ (in the absence of forcing) equal to $\pm \psi_{max}$. Hereafter, we refer to such pathlines as $O_{\pm \psi_{max}}$. 
Since $O_{\psi_{max}}$ and $O_{-\psi_{max}}$ are symmetric, we restrict our analysis to one of them only, specifically $O_{\psi_{max}}$.

Figure \ref{ChapIVFig7a} displays the linear stability property of  $O_{\psi_{max}}$ as a function of the amplitude $\epsilon$ and the frequency $\omega$ of the perturbation.
Figure \ref{ChapIVFig7a}a exhibits the residue value $R_{O_{\psi_{max}}}$ for $\epsilon \in~]0,2[$  while the frequency $\omega$ is held constant at the value $1.800$. Notice that the linear stability property depends on the value of $\epsilon$ in the following fashion.
\begin{itemize}
\item For $\epsilon \in~]0,\epsilon_{eh}\approx 0.85[$, the residue value $R_{O_{\psi_{max}}} \in~]0,1[$, indicating that  $O_{\psi_{max}}$ is elliptic. Consequently, the periodic pathline $O_{\psi_{max}}$ is surrounded by quasi-periodic pathlines forming a regular (non-mixing) island (see Fig.~\ref{ChapIVFig6a}a). 
\item For $\epsilon \in~]\epsilon_{eh},\epsilon_{he}\approx 1.15[$,
$R_{O_{\psi_{max}}} < 0$, indicating that  $O_{\psi_{max}}$ is hyperbolic. The destabilization of this key periodic pathline creates complete, or almost complete, chaotic mixing (see Figs.~\ref{ChapIVFig6a}b,~\ref{ChapIVFig8b}b).  Here the word ``complete'' (or ``almost complete'') is used in the sense that although very small non-mixing islands could appear they would be swept away by molecular diffusion due to their small size.
\item For $\epsilon \in~]\epsilon_{he},2[$,  $O_{\psi_{max}}$ is again elliptic, characterized by  surrounding regular (non-mixing) islands (see Fig.~\ref{ChapIVFig6a}c). 
\end{itemize} 

Figure \ref{ChapIVFig7a}b shows the linear stability evolution of  $O_{\psi_{max}}$ for $\omega \in~]1,3.2[$  while the amplitude $\epsilon$ is held constant at  the value $0.975$. As in the previous case, the linear stability property of $O_{\psi_{max}}$ follows  a similar scenario, as we now describe.
\begin{itemize}
\item First $O_{\psi_{max}}$ is elliptic for $\omega \in~]1,\omega_{eh}\approx 1.61[$,
indicating the presence of two large regular (non-mixing) islands (see Fig.~\ref{ChapIVFig8b}a).
\item Then, $O_{\psi_{max}}$  bifurcates from elliptic to hyperbolic at $\omega=\omega_{eh}$, and bifurcates back from hyperbolic to elliptic at $\omega=\omega_{he}\approx 1.93$. For $\omega \in ~]\omega_{eh},\omega_{he}[$, $O_{\psi_{max}}$ stays hyperbolic, indicating complete,  or at least almost complete, chaotic mixing (see Fig.~\ref{ChapIVFig8b}b).
\item Beyond the bifurcation point $\omega_{he}$,  $O_{\psi_{max}}$ is again elliptic, indicating that complete chaotic mixing does not take place due to the presence of regular (non-mixing) islands (see Fig.~\ref{ChapIVFig8b}c).   
\end{itemize}
\subsection{Existence of well spread chaotic mixing}
\begin{figure}
\begin{center}
\includegraphics*{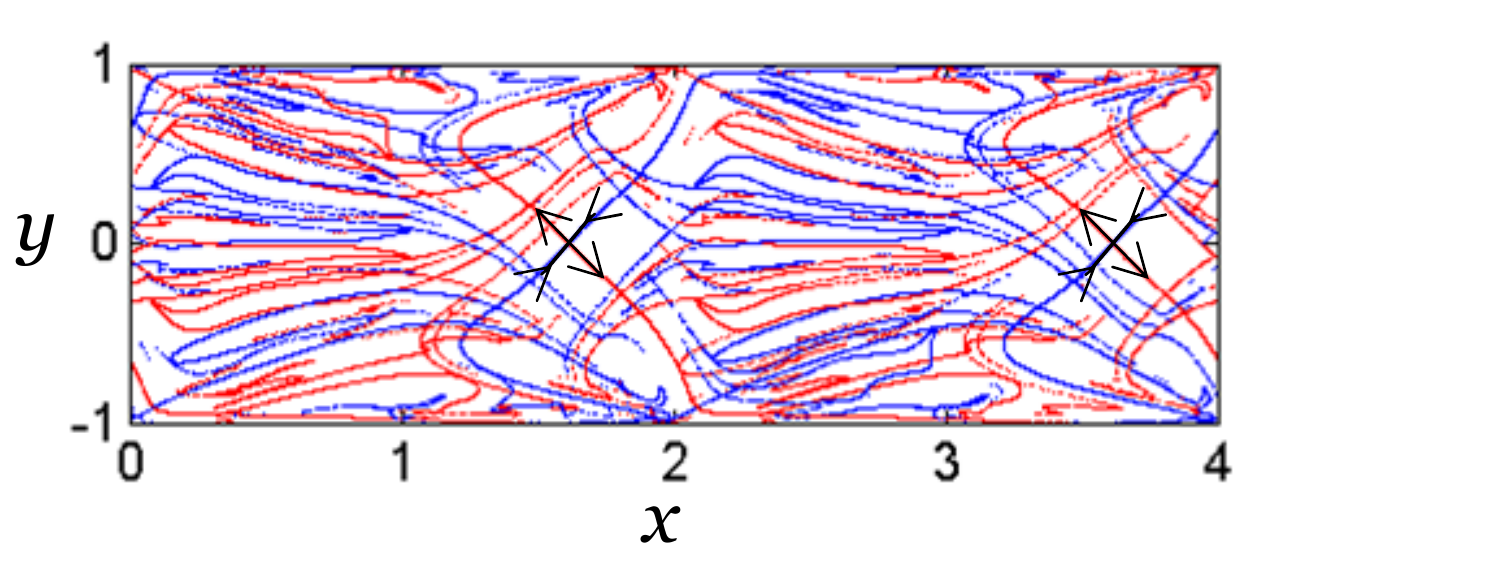}
\end{center}
\caption{Color on line. Intersections of the stable manifold, $W^s$, (blue) and the unstable manifold, $W^u$, (red) of the pair of twin periodic hyperbolic pathlines $O_{\psi_{max}}$ for the perturbation amplitude $\epsilon=0.975$ and frequency $\omega=1.800$.}
\label{figure_El_O_6}
\end{figure}

\begin{figure}
\begin{center}
\includegraphics*{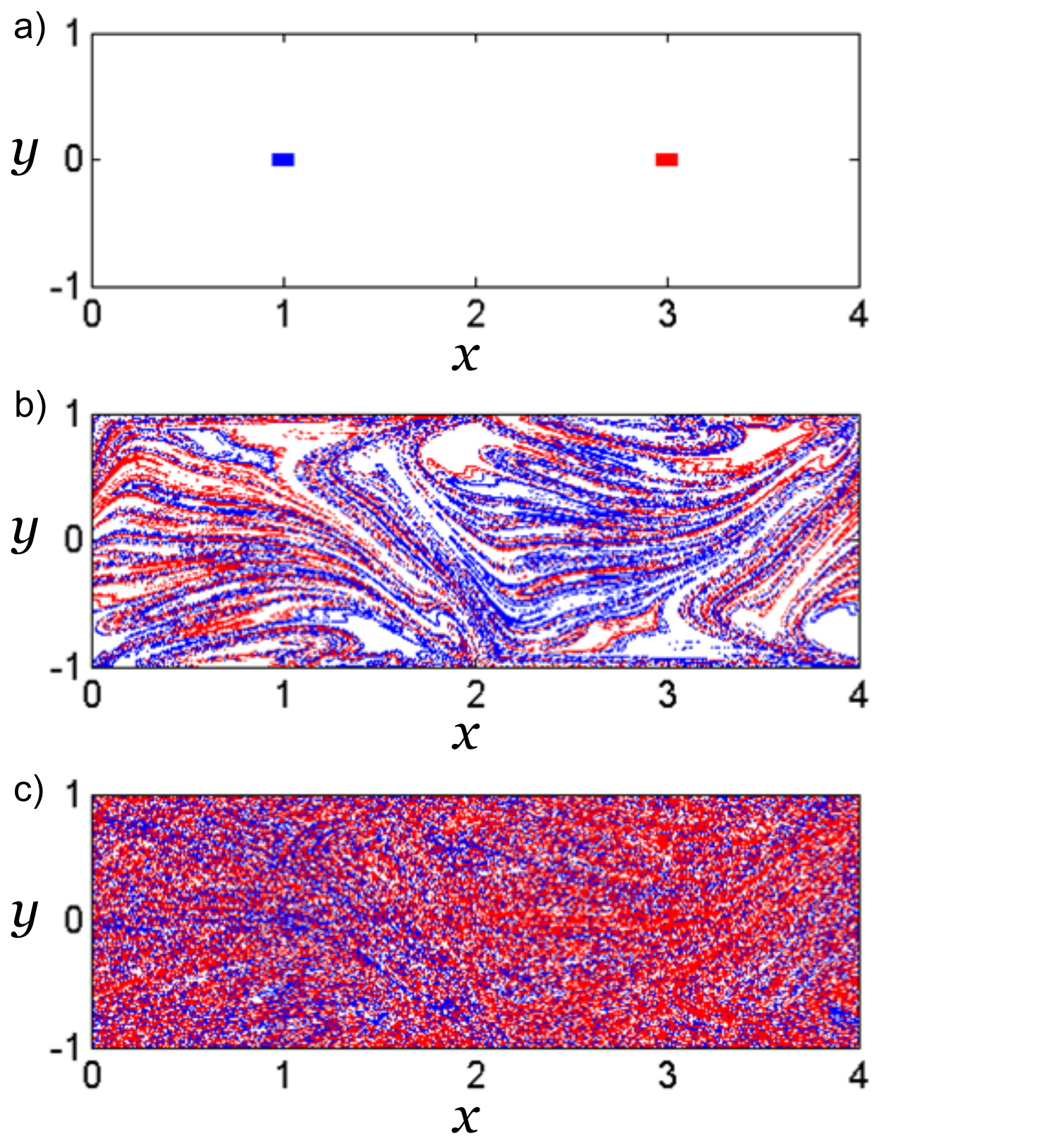}
\end{center}
\caption{Dynamics of two sets of $100,000$ particles at the times $t=0,20\pi/\omega,40\pi/\omega$ (a-c), for the perturbation amplitude $\epsilon=0.975$ and frequency $\omega=1.800$.}
\label{figure_El_O_10}
\end{figure}

The existence and spreading of chaotic mixing throughout the channel can be further studied by computing the stable and unstable manifolds of the key periodic pathlines $O_{\psi_{max}}$ in a (unstable) hyberbolic state. Figure \ref{figure_El_O_6} shows such stable and unstable manifolds $W^s$ and  $W^u$ together with their transversal intersections for  $O_{\psi_{max}}$ at the parameter values $(\epsilon=0.975, \omega=1.800)$. The transverse intersections between  $W^u$ and $W^s$ imply the existence of a {\it horseshoe map} (Smale-Birkhoff theorem  \cite{Guckenheimer:1983}), thus indicating that the system displays chaotic behavior. The horseshoe map can, in turn, be viewed as an archetypical chaotic map \cite{Ottino:1989}. Note that the lobs formed by the intersections between $W^u$ and $W^s$ are of significant sizes and are thus likely to lead to a well spread chaotic mixing region.  
\\
The well spread chaotic mixing region is illustrated in Fig.~\ref{figure_El_O_10}, where the time evolution of two sets of advected particles has been computed. Figure~\ref{figure_El_O_10} clearly shows that in less than twenty periods two sets of advected particles, each initially concentrated at the center of the two half-cells, eventually become widely mixed and well spread throughout the entire cell.

\subsection{Determination of the chaotic mixing domain in parameter space}
\begin{figure}
\begin{center}
\includegraphics*[width=12cm,height=10cm]{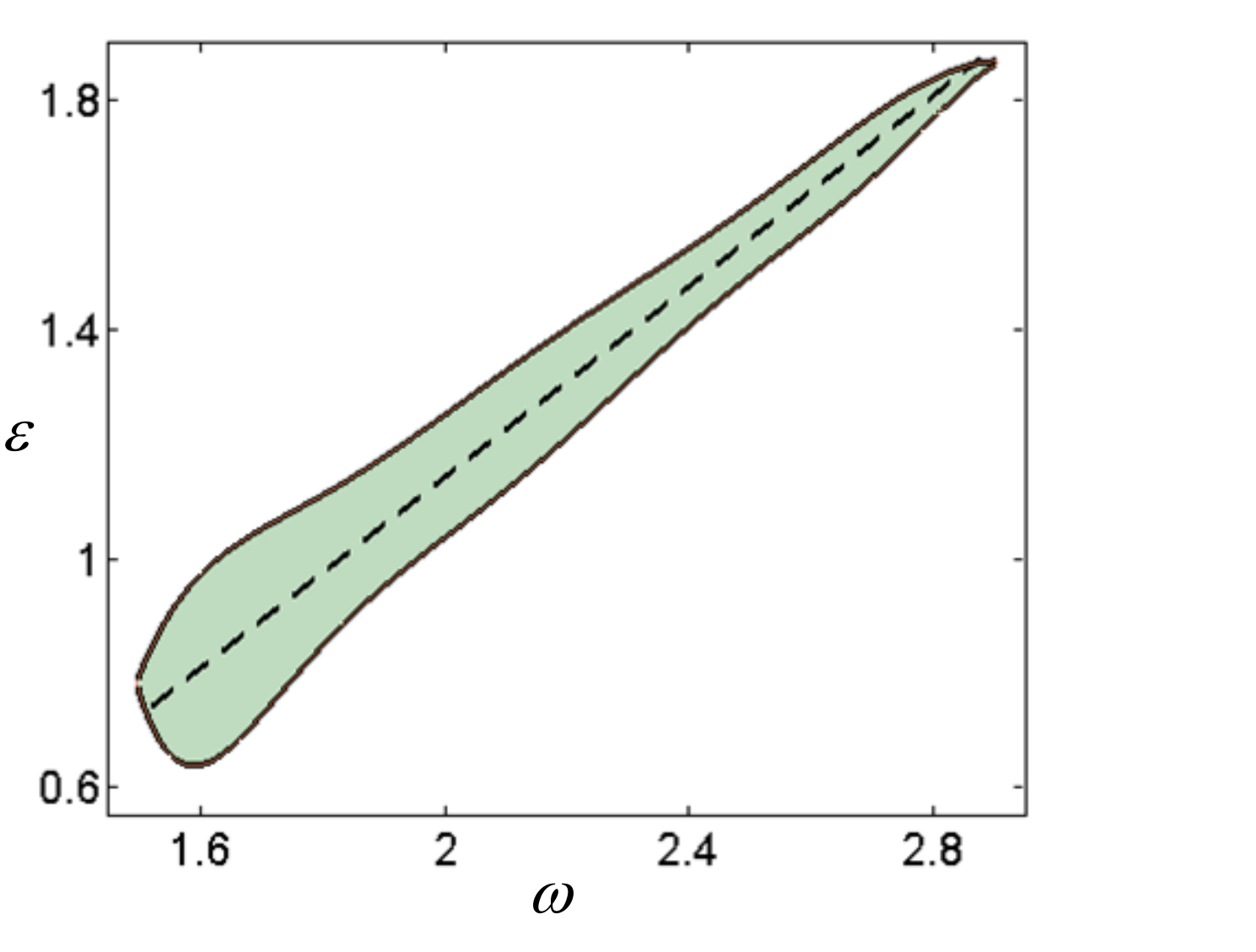}
\end{center}
\caption{
Bifurcation curve for the key periodic pathline $O_{\psi_{max}}$ (solid black line), displaying the parameter values for which $O_{\psi_{max}}$ is hyperbolic (and thus unstable), and for which complete chaotic mixing is expected (grey domain).  A linear fit of the median is also shown (dashed black line).}
\label{ChapIVFig9b}
\end{figure}

By computing the value of the residue of the periodic pathline $O_{\psi_{max}}$ as a function of both the amplitude $\epsilon$ and the frequency $\omega$, we now determine the range of parameters for which the key periodic pathline $O_{\psi_{max}}$ is hyperbolic and thus unstable. 
In Fig.~\ref{ChapIVFig9b}, the dark area corresponds to the parameter values for which $O_{\psi_{max}}$ is hyperbolic, with its boundary corresponding to the bifurcation  curve through which $O_{\psi_{max}}$ switches from being elliptic to becoming hyperbolic and vice versa (solid black line).  
From this linear stability property of $O_{\psi_{max}}$,  we deduce the region of the domain which exhibits complete (or almost complete) chaotic mixing.\\
We clearly observe that this parameter region of complete (or nearly complete) chaotic mixing stretches in a nearly linear fashion (see dashed line in Fig.~\ref{ChapIVFig9b}). 
It thus follows that we can determine a linear relationship between the amplitude of the  perturbation and its frequency in order to obtain complete chaotic mixing, which we found to be  $\epsilon= 0.83 \omega-0.53$. 
Since this parameter region stretches rather widely around such a linear median (e.g., up to $37\%$ for $\omega=1.6$), the previous relation could provide a convenient and robust guide to experimentalists desiring to select right away the optimal values of the system parameters.

\section{\label{sec:conclu}Conclusion}
In this paper, we have presented an efficient strategy for locating
the complete, or almost complete, chaotic regime in parameter
space, i.e., as the amplitude and frequency of the perturbation is varied.
In contrast with the common trial and error methods requiring the integration
of numerous trajectories over long periods of time, the strategy presented here 
focused on a few invariant trajectories, i.e., key periodic pathlines over a short period of time. Specifically, we have studied the evolution in time of the linear stability
of these key periodic pathlines as a function of the system
parameters and predicted the spreading of chaotic mixing as such pathlines destabilize. We have also shown both qualitatively (using Poincar\'e sections and
stable and unstable manifolds intersections) and quantitatively (using a finite time Lyapunov map and a mixing index) that such a local approach leads to an accurate prediction of complete (or almost complete) chaotic mixing, while being more efficient than the trial and error methods.
Finally, for the electroosmotic flow considered in this paper, we have determined the sub-domain of parameter values producing complete (or almost complete) chaotic mixing. Given the quasi-linear shape of the sub-domain of parameters for which complete (or almost complete) mixing takes place, we have also proposed a linear relation between the parameters which should be useful as a guide to experimentalists who can then readily adapt the system parameters to optimal values.  
\bibliographystyle{pf}
 \bibliography{MyBiblio}

\end{document}